\documentclass[12pt]{article}
\usepackage{graphicx}
 \usepackage{amssymb}
 \usepackage{amsmath}
\usepackage{cite}
 \usepackage{geometry}               
\newcommand {\slsh} [1] {\not{\hbox{\kern-2pt${#1}$}}}

\newcommand {\beq} {\begin{equation}}
\newcommand {\eeq} {\end{equation}}
  \newcommand {\ber}{\begin{eqnarray*}}
  \newcommand {\eer} {\end{eqnarray*}}
\newcommand {\beqn}{\begin{eqnarray}}
  \newcommand {\eeqn} {\end{eqnarray}}

\begin{document}
\begin{titlepage}
\begin{flushright}
{FTPI-MINN-05/21\\
UMN-TH-2406/05\\
}
\end{flushright}
\vskip 0.8cm

\centerline{{\Large \bf Highly Excited 
Hadrons in QCD and Beyond}\,\footnote{Based on 
invited talks delivered at the {\sl First Workshop on Quark-Hadron 
Duality and the Transition to pQCD}, Laboratori Nazionali di Frascati, 
Italy, June 6--8, 2005, and the {\sl Workshop 
on Highly Excited Hadrons}, ECT$^*$ in Trento, Italy, July 4--9,  2005.}}
\vskip 1cm
\centerline{\large  M. Shifman  
}
\vskip 0.1cm
\begin{center}
{\em  William I. Fine Theoretical Physics Institute, 
University
of Minnesota, Minneapolis, MN 55455, USA }
\end{center}

\vspace{1.5cm}

\begin{abstract}

\vspace{.3cm}

I discuss two issues related to high ``radial" excitations
which attracted much attention recently:
(i) chiral symmetry restoration in excited mesons and baryons,
and  (ii) universality of the
$\rho$-meson coupling  in QCD and AdS/QCD. New results
are reported and a
curious relation between an AdS/QCD formula and 1977 
Migdal's proposal is noted.

 \end{abstract}

\end{titlepage}

\section{Introduction}

A renewed interest to highly excited mesons in QCD is explained by
at least two reasons. 

First, the gravity/gauge correspondence, originally established for 
conformal field theories on the gauge side, is being aggressively expanded to include
closer relatives of QCD, with the intention to get a long-awaited
theoretical control over QCD proper. 
The present-day ``bottom-up" approach  is as follows:
one starts from QCD and attempts to guess its
five-dimensional holographic dual. In this way, various
holographic descriptions of QCD-like theories 
emerge and their consequences are being scrutinized with the purpose of finding
the fittest model. 

This approach goes under the name 
``AdS/QCD."  Since the limit $N=\infty$ is inherent to
AdS/QCD, the meson widths vanish, and one can  unambiguously define  
masses and other static characteristics of excited states. Explorations 
of this type were reported in 
Refs. \cite{K,1,Kr,Sakai,Kruc,B,H,E,Bur,Kleba,2,3,4,5,drp,Kup,Kirsch}. In  refined
holographic descriptions expected to emerge in the future
one hopes to get asymptotically linear primary and daughter Regge trajectories, 
and obtain residues and other parameters and regularities pertinent
to hadronic physics, and demonstrate their compatibility with experiment ---
a goal which has not yet been achieved.

The recently suggested orientifold large-$N$ expansion \cite{ASV}
which is complementary to the 't Hooft one \cite{thooft}, provides
another framework which can be used for studying 
excited states.

The second reason is a clear-cut  demonstration that the chiral symmetry 
of the QCD Lagrangian is  empirically restored
in  excited mesons and baryons, due to 
Glozman and collaborators \cite{Gloz0,Gloz1,Gloz2,Gloz3}.
As is well-known from the early days of QCD, highly excited hadrons
can be described quasiclassically (see e.g. \cite{NSVZ}; recapitulated recently 
in Ref. \cite{Gloz4} at the qualitative level). The quasiclassical 
description implies, in particular, asymptotically linear Regge trajectories. 
Needless to say, it also implies
linear realization of all symmetries of the QCD Lagrangian at high energies,
in particular, in high radial excitations.
Indeed, in such states the valent quarks on average have high
energies --- high compared to $\Lambda_{\rm QCD}$ ---
and, thus, can be treated quasiclassically so that effects 
due to the quark condensate are inessential and can be neglected.

Nevertheless, the fact that excited states do exhibit 
the full linearly-realized  chiral symmetry of QCD
seemingly caught some theorists by surprise, probably, due to  an
over-concentration
on the low-lying states for which the chiral symmetry is
broken. (It would be more exact to say that
the axial SU$(N_f)$ where $N_f$  is the number
of   massless flavors is
realized non-linearly, in the Goldstone mode.)
According to  \cite{Gloz0,Gloz1,Gloz2,Gloz3}, the 
symmetry is largely restored 
already for the first radial excitations; for instance, three excited pions and
one excited scalar-isoscalar meson form an almost degenerate
dimension-4 representation of 
SU(2)$\times$SU(2)$\sim$O(4).\footnote{I assume for 
simplicity that there are two massless flavors, ignoring the strange
quark. The chiral U(1) also gets restored as a valid flavor symmetry:
the U(1) chiral anomaly dies off in the limit of large
number of colors.
Even if the number of colors $N$ is kept fixed, at large energies (i.e. high $n$),
where perturbation theory becomes, in a sense, applicable,
the axial U(1) gets restored since one can always
redefine the quark U(1) current by adding a Chern-Simons current
in such a way that the U(1) charge is conserved in perturbation theory.
Restoration of the axial U(1) in radially excited mesons
is clearly seen in experiment, cf. e.g. \cite{Gloz2}.}

The question is what can be said quantitatively on the
rate of the symmetry restoration. In other words, what is the 
$n$-dependence ($n$ is the excitation number) of  the splittings 
$\delta M_n$ of the
radially  excited states from one and the same representation  of 
SU(2)$\times$SU(2)? 

In the first part of my talk  I suggest some  natural 
``pedestrian"  estimates  of the
rate of the chiral symmetry restoration. In the second part
I will focus on an aspect of AdS/QCD which was recently discussed
in \cite{4,5}: implementation of the vector meson dominance (VMD)
and universality of the $\rho\, HH$ coupling. AdS/QCD-based results
will be confronted with QCD expectations. 

\section{Chiral symmetry restoration: generalities}

If both vector  and axial SU(2)'s are linearly realized, hadronic states 
must fall into degenerate multiplets of the full chiral symmetry.
The degeneracy is lifted by an SU(2)$\times$SU(2)$\to$SU(2) 
breaking that dies off
with energy (or the excitation number, which is the same).
I will first briefly review an appropriate representation theory \cite{Gloz2}.
My next task will be  an estimate of $\delta M_n$ versus $n$ at $n\gg 1$
where  $\delta M_n$ is the mass difference in particular
chiral representations, for instance, the mass difference of the
scalar--pseudoscalar mesons or the vector--axial-vector ones.

It is clear that to define highly excited states {\em per se}
we need large $N$. In tricolor QCD, as we will  shortly see,
resonance widths rapidly grow with the excitation number; as a result,  for 
large $n$ the very separation of  mesons 
from continuum becomes impossible, and the question of mass splittings 
and other similar questions
cannot be addressed.
{\em I will systematically exploit the large}-$N$ {\em limit}.
I will consider the massless quark sector consisting of two  flavors.
Extension to three flavors is rather straightforward.

The large-$N$ limit sets an appropriate theoretical framework
for consideration of excited mesons. As for excited baryons,
there is no reason for their  widths to be suppressed at
$N\to\infty$. Therefore, theoretical analysis of highly excited
baryonic states becomes problematic, as at $n\gg 1$ they should form a continuum. 
I will limit my discussion to mesonic states.
(Empirically, isolation  of  excited baryon resonances
from existing hadronic data  seems possible, and data seem to indicate  
restoration of the full chiral symmetry
for excited baryons \cite{Gloz0,Gloz1,Gloz2,Gloz3}. Representations
of unbroken SU(2)$\times$SU(2) for baryons had been 
studied long ago, even before the advent of QCD \cite{jido}.)

Analysis of the chiral symmetry is more conveniently performed in terms of the
Weyl rather than Dirac spinors. Each Dirac spinor is a
pair of two Weyl ones. We can take them both to be left-handed,
$\chi$ and $\eta$, the first being triplet with respect to SU(3)$_{\rm color}$,
the second antitriplet.\footnote{If the gauge group is extended to SU($N$),
with respect to SU($N$)$_{\rm color}$ the field $\chi$ transform as $N$,
while $\eta$ as $\overline N$.}
To form color singlets we convolute
$\chi\eta$ or, alternatively, $\bar\chi\chi$, $\bar\eta\eta$.
The Dirac spinor $\Psi$ combines one left-handed and one
right-handed spinor, $\Psi\sim \{\chi_\alpha ,\,\, \bar\eta_{\dot\beta}\}$
where $\alpha$ and $\dot\beta$ are spinorial indices.
Each chiral spinor carries a flavor SU(2) index. Since there are two 
linearly realized SU(2)'s,
there are two flavor SU(2) indices,
$\chi^k$ and $\eta^{\dot a}$ ($k,\dot a, =1,2$)
which are independent.
We can call them ``left" and ``right" isospin. Conventional isospin
entangles ``left" and ``right" isospins.

The exact conserved quantum
numbers of QCD, namely, conventional isospin, 
total angular momentum and parity,
do not always completely specify the full structure of a
quark-antiquark meson, as we will see shortly. Distinct patterns
of ``left" and ``right" isospin additions can lead
to distinct mesons having the same
conventional isospin, 
total angular momentum and parity.

Let us now discuss the simplest representations.
The scalar (pseudoscalar) mesons are of the type
\beq
\chi\eta \pm \bar\chi\bar\eta\,.
\eeq
Its Lorentz structure is $(1/2,0)\times (1/2,0)\to (0,0)$
and $(0,1/2)\times (0,1/2)\to (0,0)$. The isospin structure is
$(1/2,1/2)$. In terms of conventional isospin
this is a triplet plus a singlet. In terms of 
SU(2)$\times$SU(2)$\sim$O(4) we have two four-dimensional chiral representations:
The first includes scalar isoscalar plus pseudoscalar isovector, the second
pseudoscalar isoscalar plus 
scalar isovector.

In fact, the symmetry that gets restored is 
higher than just SU(2)$\times$SU(2)$\sim$O(4). Indeed, at $N\to\infty$ the two-point functions
of the currents $\bar\Psi\Psi$ and $\bar\Psi \tau^a \Psi$ are degenerate
(here $\tau^a$ are the conventional isospin generators, $a=1,2,3$), 
since the quark-gluon mixing that can occur in the isoscalar ---  but not isovector 
---  channel is suppressed by $1/N$ and, thus, can be neglected.
The above degeneracy is in one-to-one correspondence with the fact
that the full flavor symmetry of the QCD Lagrangian is 
U(1)$_V\times$U(1)$_A\times$SU(2)$\times$SU(2).
The vector U(1), the baryon charge, plays no role in the meson
sector. The linear realization of U(1)$_A\times$SU(2)$\times$SU(2)
implies that the
two four-dimensional representations of 
SU(2)$\times$SU(2) are combined in one irreducible
eight-dimensional representation
of U(1)$_A\times$SU(2)$\times$SU(2).

If we pass
to nonvanishing angular
momenta, we observe that
the vector and axial-vector mesons can be of 
two types,\footnote{This fact was, of course, known to 
QCD practitioners from the very beginning, and was repeatedly exploited in QCD and in models
in various contexts, e.g. \cite{Ch}.}
\beqn
\label{ju}
&&\bar\chi_{\dot\alpha}\chi_\alpha \pm 
\bar\eta_{\dot\alpha}\eta_\alpha \,,\\[2mm]
\label{an}
&& \chi_{\{\alpha}\eta_{\beta\}} \pm \mbox{h.c.}\,,
\eeqn
where the braces denote symmetrization. The first one is Lorentz $(1/2,1/2)$,
the second $(1,0)+(0,1)$.
In terms of the ``left" and ``right" isospins, it is
the other way around. The state $\bar\chi_{\dot\alpha}\chi_\alpha$ has isospin 
$(\overline{1/2},0)\times (1/2,0)\to (1,0) +(0,0)$. 
It is a triplet plus a singlet with respect to the
conventional isospin. Taking into account the second term in (\ref{ju}),
we have a vector isovector plus an axial-vector isovector
plus two isosinglets. At large $N$ they all, taken together,
form an eight-dimensional representation.
The state (\ref{an}) forms SU(2)$\times$SU(2) quadruplets  $(1/2,0)\times
(0,1/2)$: a vector isoscalar plus an axial-vector isovector, and vice versa,
a vector isovector plus an axial-vector isoscalar. Again, both quadruplets are combined
into an eight-dimensional representation at $N\to\infty$.

A physical ground-state $\rho$ meson which is roughly
an equal mixture of (\ref{ju}) and (\ref{an}) 
is ``polygamous" and has two distinct chiral partners
\cite{Gloz2}: an axial-vector isovector and an axial-vector isoscalar.

\section{The rate of the chiral symmetry restoration}
\label{trotcsr}

For simplicity I will discuss the mass splittings of 
scalar versus pseudoscalar excited mesons, produced from the vacuum by the
operators $\bar\Psi\Psi$ and $i\bar\Psi\gamma_5\Psi$, respectively.
As was mentioned, at $N\to\infty$ their isotopic structure
is inessential.
The mass splittings in other chiral multiplets must have
the same $n$ dependence.

The chiral symmetry is broken by the quark mass term, 
which I will put to zero. Then the local order parameter
representing the chiral symmetry breaking is
$\langle\bar\Psi\Psi\rangle$. Unfortunately, it is rather hard 
to express the mass splittings in the individual multiplets in terms
of this local parameter. Generally speaking, the fact
that its mass dimension is 3 tells us that various chiral
symmetry violating  effects
will die off as inverse powers of $M_n$ (i.e. as positive powers of 
$n^{-1/2}$).  Today's level of command of QCD  does 
not allow us to unambiguously predict the laws of fall off of 
the  chiral symmetry violating  effects in terms 
of $\langle\bar\Psi\Psi\rangle$.
In some instances a minimal rate  can be estimated, however.
In some instances there are reasons to believe that 
the actual rate may be close to the minimal one.

\subsection{On quasiclassical arguments}

As a warm-up exercise I will derive textbook results:
equidistant spacing of radially excited mesons and 
$n$ independence of $\Gamma_n/M_n$ where $\Gamma_n$
is the total width of the 
$n$-th excitation. In the following simple estimates I will try 
to be as straightforward as possible,
omitting inessential numerical constants 
and assuming that the only mass dimension is provided
by $\Lambda_{\rm QCD}\equiv \Lambda$.
In this ``reference frame"
the string tension is $\Lambda^2$, while the $\rho$-meson mass
is $\Lambda$.

When a highly excited meson state (say, $\rho_n$)
is created by a local source (vector current), it can
be considered, quasiclassically, as a pair of   free
ultrarelativistic
quarks; each of them with   $E=p=M_n/2$. These quarks are 
produced
at the origin, and then fly back-to-back, eventually creating  a flux 
tube of the chromoelectric field. Since the tension of the flux tube is
$\sigma \sim \Lambda^2$,
the length of the tube 
is 
\beq
L\sim 
\frac{M_n}{\Lambda^2}\,.
\label{one}
\eeq
Using the quasiclassical quantization condition
\beq
\int p\, dx \sim p \, L \sim \frac{M_n^2}{\Lambda^2} \sim n
\label{two}
\eeq
we immediately arrive at
\beq
M_n^2 \sim \Lambda^2 \, n\,.
\label{three}
\eeq
(Let me parenthetically note that asymptotically linear $n$ dependence of $M_n^2$
can be analytically obtained   in the two-dimensional  't Hooft model
\cite{TM, TM1} where linear confinement is built in.)

Let us now discuss total decay widths of high radial excitations.
The decay 
probability (per unit time) is determined, to order $1/N$, 
by the probability of producing an extra quark-antiquark pair. Since 
the pair creation can happen anywhere inside the flux tube, the
probability must be proportional to $L$. 
As a result \cite{Nussinov},
\beq
\Gamma_n\sim \frac{1}{N}\,  L\Lambda^2 =\frac{B}{N}\, M_n\,,
\label{estgam}
\eeq
where $B$ is a dimensionless coefficient independent
of $N$ and $n$, see Eq.~(\ref{one}).

Thus, the width of the $n$-th excited state
is proportional to its mass which, in turn, 
is proportional to $\sqrt{n}$.
The square root formula for $\Gamma_n$ was numerically 
confirmed \cite{BS} in the   't Hooft model. 
It is curious that both, in actual QCD and in two
dimensions, $B\sim 0.5$.

Equation (\ref{estgam}) demonstrates that the limits
$N\to\infty$ and $n\to\infty$ are not commutative.
We must first send $N$ to infinity, and only then can we consider high  
radial excitations.

Asymptotic linearity of the Regge trajectories (Eq. (\ref{three}) at $n\gg 1$)
must emerge in any sensible string-theory-based description of
QCD. As discussed in detail by Schreiber \cite{Schreiber},
this is indeed the case in the picture of mesons as open strings
ending on a probe D brane in an appropriate background.
In the same work,
using an open string analog of the well-known Witten's argument, Shreiber shows
\cite{Schreiber} that treating radial excitation of low-spin mesons
as fluctuations of the probe D branes one obtains, generally speaking, a wrong
behavior, $M_n^2 \sim \Lambda^2 \, n^2$.  (This remark which pre-emps the beginning of Sect.~\ref{adsvqcd} will be explained there in more detail.) 

\subsection{Analyzing chiral pairs: the slowest fall off of $\delta M_n$}

For definiteness I will focus on scalar--pseudoscalar mesons.
Semi-quantitative results to be derived below are 
straightforward and general.
The only ``serious" formula I will use is that for the Euler function,
\beq
\psi (z) =-\gamma -\frac{1}{z} +\sum_{n=1}^\infty
\left(\frac{1}{n} -\frac{1}{z+n}
\right)\,.
\label{four}
\eeq
At large positive $z$
\beq
\psi (z)\to \ln z -\frac{1}{2z} +O(z^{-2})\,.
\label{five}
\eeq
The two-point correlators we will deal with are defined as
\beq
\Pi (q) = i\int\,d^4xe^{i\,qx}\,  \langle T\{ J(x),\, J(0)\rangle \,,
\label{six}
\eeq
where
\beq
J =\bar\Psi \Psi \,\,\,\mbox{and} \,\,\, J_5 =i\, \bar\Psi \gamma^5 \Psi   
\label{seven}
\eeq
for scalars and pseudoscalars, respectively (to be denoted as 
as $\Pi$ and $\Pi_5$). 
I will consider flavor-nonsinglet channels.
Then
\beq
\Pi (Q) - \Pi_5(Q) =\sum_n \left(\frac{f_n}{Q^2 +M_n^2} -
\frac{\tilde f_n}{Q^2 +\tilde M_n^2}
\right) \,,
\label{scps}
\eeq
where the untilded quantities refer to the scalar channel
while those with tildes to the pseudoscalar one.

The operator product expansion (OPE) for $\Pi (Q) - \Pi_5(Q)$
at large Euclidean $Q^2$ was built long ago \cite{SVZ}.
In conjunction with the large-$N$ limit
which justifies factorization of the four-quark operators it
implies
\beq
\Pi (Q) - \Pi_5(Q) \sim \frac{1}{N}\, 
\frac{\langle\bar\Psi\Psi\rangle^2}{Q^4}\,,\qquad Q^2 \to \infty\,,
\label{mat}
\eeq
(modulo possible logarithms).
Now, the residues $f_n$ are positive numbers of dimension $\Lambda^4$.
They can be normalized from the relation
\beq
\Pi (Q) \sim N Q^2 \ln Q^2\,,\qquad Q^2 \to \infty\,.
\eeq
In fact, it is easy to show that the equidistant spectrum
(\ref{three}) combined with Eq.~(\ref{four}) lead to   
\beq
f_n \sim N \Lambda^2 M_n^2\,.
\label{scps1}
\eeq

\vspace{2mm}

Now we are in position to estimate the splittings.
As was explained above, one expects that asymptotically, at large $n$,
\beq
| \delta f_n |\ll f_n\,,\qquad | \delta M^2_n |\ll M^2_n\,,
\label{eight}
\eeq
where
\beq
\delta f_n =f_n - \tilde f_n\,,\qquad \delta M^2_n =M^2_n -\tilde M^2_n\,.
\eeq
The scalar-pseudoscalar difference in Eq.~(\ref{scps}) depends
on $\delta f_n$ and $\delta M^2_n$. We first set
$\delta f_n =0$, i.e. assume perfectly degenerate residues.
(Shortly this degeneracy will be lifted, of course.)
Then, taking account of (\ref{scps1}), we get
the following sum-over-resonances representation:
\beqn
\Pi (Q) - \Pi_5(Q) 
&=& N\Lambda^2 \, Q^2\, \sum_n \frac{\delta M^2_n }{\left( Q^2 +M_n^2\right)^2}
\nonumber\\[3mm]
&\to&
N\Lambda^2 \, Q^2\,\frac{\partial}{\partial Q^2}\sum_n \frac{\delta M^2_n }{ Q^2 +M_n^2 }\,.
\label{mat1}
\eeqn

To evaluate the convergence of $\delta M_n^2$ to zero,
the last expression must be matched with the asymptotic formula
(\ref{mat}) at $Q^2\to\infty$.
Needless to say, matching an infinite sum to a single OPE term
one cannot expect to get a unique solution for $\delta M_n^2$.
In fact, what we are after, is the {\em slowest}  pattern 
of the chiral symmetry restoration still compatible with the
OPE.\,\footnote{As we will see in Sect.~\ref{adsvqcd},
the advent of holographic ideas in QCD definitely 
revived interest in combining OPE with Regge-trajectory-based
constructions. For instance, 
attempts to 
reconcile OPE in the chirality-breaking two-point function
$\langle V+A,\,\, V-A\rangle$ with strictly linear Regge trajectories 
were reported  in \cite{BBB}. Of course, no 
reconciliation can be achieved under the assumption of exact linearity.
Similar arguments were later 
used in \cite{AAA} to address the issue of 
asymptotic deviations of the Regge trajectories from
linearity.}

Even if $\delta M^2_n$ falls very fast at large
$n$, but the sign of $\delta M^2_n$ is $n$-independent,
there is no matching. Indeed, 
Eq.~(\ref{mat1}) implies then $1/Q^2$ rather than $1/Q^4$
behavior. The only way to enforce the $1/Q^4$ asymptotics is to assume that
$\delta M^2_n$ is a sign alternating function of $n$, say,
\beq
\raisebox{-1.6ex}{\rule{0mm}{10mm}}
\delta M^2_{2k} >0\,,\qquad \delta M^2_{2k+1} < 0\,,
\qquad |\delta M^2_{2k}| = |\delta M^2_{2k+1}|\,.
\eeq
Let us show that the slowest possible
decrease of $| \delta M^2_n | $ versus $n$ is 
\beq
\raisebox{-1.6ex}{\rule{0mm}{10mm}}
M^2_n \delta M^2_n =\mbox {sign alternating const}\,.
\label{mat2}
\eeq

Equation (\ref{mat}) implies that $\delta \Pi (Q^2)$ is analytic in the complex plane,
with a $Q^{-4}$ singularity at the origin. This implies, in turn, that the sum
\beq
\sum_n \, \delta M_n^2
\label{convsum}
\eeq
is convergent and  vanishes. The slowest-rate solution is
$\delta M_n^2 \sim (-1)^n\, n^{-1}$, which is the same as (\ref{mat2}).
Under the condition (\ref{mat2})
the right-hand side of (\ref{mat1}) becomes
\beqn
&& N\Lambda^6 \, Q^2\,\frac{\partial}{\partial Q^2}\sum_k
\left\{  \frac{1}{M^2_{2k}\left( Q^2 +M^2_{2k}\right)} -
 \frac{1}{M^2_{2k+1}\left( Q^2 +M^2_{2k+1}\right)} 
\right\}\nonumber\\[4mm]
&&\to  N\Lambda^4\, Q^2\,\frac{\partial}{\partial Q^2}\frac{1}{Q^2}
\ln \frac{Q^2+\Lambda^2}{Q^2}\to N\Lambda^6 \frac{1}{Q^4}\,,
\label{ib}
\eeqn
{\em quod erat demonstrandum. } 
Here $\Lambda^6$ must be identified with 
$N^{-2}\langle\bar\Psi\Psi\rangle^2$.

\vspace{2mm}

\begin{figure}[h]
 \centerline{\includegraphics[width=2.5in]{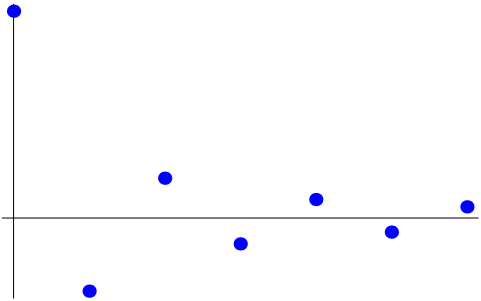}}
 \caption{\small $\delta M_n$ versus $n$, in arbitrary units.}
 \label{tr1}
 \end{figure}

Thus, $\delta M_n$ is sign alternating and falls off as
\beq
| \delta M_n |   \sim  \frac{\Lambda}{n^{3/2}}
\label{zib}
\eeq
or {\em faster}. Whether or not it actually falls off faster than $n^{-3/2}$
depends on the behavior of higher order terms in OPE
for $\Pi (Q) - \Pi_5(Q)$. Next-to-nothing can be said on this at the moment.
Even with the slowest possible regime (\ref{zib})
the rate of the chiral symmetry restoration is pretty fast, see
Fig.~\ref{tr1}.

\subsection{The impact of \boldmath{$\delta f_n\neq 0$}}

Now, let us relax the (unrealistic) assumption 
$\delta f_n = 0$. Again, our task is to find the slowest fall off
compatible with both the OPE and resonance representations.
If both $\delta f_n \neq  0$ and $\delta M_n \neq  0$,
the difference $\Pi (Q) - \Pi_5(Q)$ acquires an additional term
\beq
\Pi (Q) - \Pi_5(Q) =\sum_n  \frac{\delta f_n}{Q^2 +M_n^2} \,.
\label{scps3}
\eeq
It is not difficult to see that, barring an unlikely and subtle conspiracy between
$\delta f_n  $ and $\delta M_n $, the previous estimate for
$\delta M_n $ stays intact, and, in addition, one gets an
estimate for $\delta f_n  $.
Copying the consideration above one finds that
$\delta f_n$ must be sign-alternating, and the fall-off
regime of $|\delta f_n |$ must be $N\Lambda^4/n$ or faster,
so that 
\beq
\raisebox{-1.6ex}{\rule{0mm}{10mm}}
|\delta f_n|/f_n \sim 1/n^2\,.
\label{piz}
\eeq
Here I used the fact that $f_n$ grows linearly with $n$, see Eq.~(\ref{scps1}).

\subsection{Excited  $\rho$ mesons of two kinds}

In Ref.~\cite{Gloz2} it was noted that, if the chiral symmetry is restored
at high $n$, two distinct varieties of excited $\rho$ mesons must exist.
Let us define $\rho$ meson as a $J^{PC}=1^{--}$ quark-antiquark state
with isospin 1. Then, such mesons are produced from the vacuum
by the following two currents\,\footnote{The current
(\ref{artwo}) also produces $1^+$ mesons;
this is irrelevant for what follows.}
which belong to two distinct
chiral multiplets (see Eqs. (\ref{ju}) and (\ref{an})):
\beqn
&&\bar\Psi\vec\tau\gamma_\mu \Psi\,,
\label{arone}\\[2mm]
&&
\bar\Psi\vec\tau
\sigma_{\alpha\beta} \Psi \,.
\label{artwo}
\eeqn

\begin{figure}[h]
 \centerline{\includegraphics[width=2.5in]{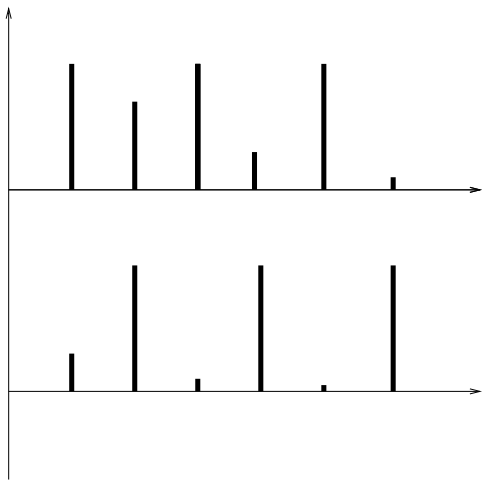}}
 \caption{\small Spectral densities in the channels corresponding to the currents (\ref{arone}) and (\ref{artwo}). Contributions due to 
 $\rho_n$ of the ``wrong kind" die off as $1/n^3$ at large $n$.}
 \label{poezd1}
 \end{figure}

Were the chiral symmetry exact, the above two currents would not mix.
The ground state $\rho$ meson would be coupled to the first current
and would have vanishing residue in the second.
There will be two distinct types of the $J^{PC}=1^{--}$ mesons.
In actuality, they do mix, however;
the $\rho$ mesons show up in both currents, see Fig.~\ref{poezd1}.
The $J^{PC}=1^{--}$ excited states 
of the first kind couple predominantly to the first current while
those of the second kind to the second current.
I want to  evaluate the coupling of $\rho_n$ of a given kind 
to the ``wrong" 
current (as a function of $n$).

To this end we will consider a ``mixed" correlation function
\beq
\left\langle\bar\Psi\vec\tau\gamma_\mu \Psi\,,\,\,\bar\Psi\vec\tau
\sigma_{\alpha\beta} \Psi\right\rangle_q
\sim
\left(
g_{\alpha\mu}\,q_\beta - g_{\beta\mu}\, q_\alpha
\right)
\,\langle\bar\Psi\Psi\rangle\,\frac{1}{q^2}
\label{clione}
\eeq
modulo logs of $Q^2$.
Saturating this expansion by resonances and defining
\beq
\langle {\rm vac} |\bar\Psi\vec\tau
\sigma_{\alpha\beta} \Psi|\rho_n\rangle=
\left(
g_{\alpha\mu}\,p_\beta - g_{\beta\mu}\, p_\alpha
\right)f_n\,\varepsilon_\mu^a
\label{clitwo}
\eeq
we get, say, for the resonances ``strongly" coupled to
$\Psi\vec\tau\gamma_\mu \Psi$
\beq
\sum_n \frac{M_n^2}{g_n} \, f_n \sim \langle\bar\Psi\Psi\rangle\,,
\label{clithree}
\eeq
implying, in turn, that
\beq
f_n\sim \frac{\Lambda}{n^{3/2}}
\label{clifour}
\eeq
or faster. Thus, barring conspiracies,
the contribution, of an excited $\rho$
of the ``wrong" chiral structure in the {\em diagonal} two-point function
of two vector currents $\bar\Psi\vec\tau\gamma_\mu \Psi$
scales as
$f_n^2 \sim 1/n^3$, to be compared with that of the ``right" chiral
structure,
$M_n^2/g_n^2\sim n^0$. The relative suppression is $1/n^3$ or faster.

One can arrive at the very same statement
using  a slightly different argument.
The wrong-chirality $\rho_n$'s in the {\em diagonal} two-point function
of two vector currents $\bar\Psi\vec\tau\gamma_\mu \Psi$
must be associated with the operator $\langle\bar\Psi\Psi\rangle^2$.
Therefore, the sum
$\sum_n f_n^2\, M_n^4$ must converge, again leading to the estimate
$f_n^2 \sim 1/n^3$ or faster.

\section{AdS/QCD versus QCD}
\label{adsvqcd}

As I have already mentioned, AdS/CFT correspondence
\cite{AdSCFT} inspired a general  
search for five-dimensional holographic duals of QCD. 
This trend flourished in the last few years, e.g. 
\cite{K,1,Kr,Sakai,Kruc,B,H,E,Bur,Kleba,2,3,4,5,drp,Kup,Kirsch}. Generic holographic models of QCD 
describe gauge theories that typically (but not always!)
have color confinement and chiral symmetry breaking as built-in features,
and are dual to a string theory on a weakly-curved background.
One should remember that they describe strong coupling 
theory --- i.e. asymptotic freedom is not properly incorporated. 
In the majority of the holographic duals discussed so far
$M_n\sim n$ rather than $\sqrt n$ required by
linearity of the  Regge trajectories. In some more contrived
holographic descriptions (e.g. \cite{Kup}) asymptotically linear Regge 
trajectories do emerge, however. 

In discussing flavor physics in AdS/QCD one should keep in mind that
at present there are two methods of introducing
dynamical quarks in the  fundamental  representation into the
gravity/string duals of QCD: (i) the so-called flavor probes (for $N_f\ll N$); (ii)
fully back-reacted backgrounds that incorporate flavor --- in this case, obviously,
there are no restrictions on $N_f$. The latter direction is not yet 
sufficiently developed. Burrington et al. \cite{Bur} 
suggested a background based on D3/D7 system, which seems singular,
however. This construction has some undesirable
features, serious drawbacks, which cannot be 
discussed here. Klebanov and Maldacena incorporated
flavor \cite{Kleba} in a conformal theory claimed to be  dual to the infrared fixed
point of ${\cal N}=1$ SQCD, a theory which is in no way close to QCD.
This seems to exhaust the list of developments in this direction.

The
probe approach pioneered by Karch and Katz \cite{K}
is way more advanced. In the original work \cite{K}
Karch and Katz added D7 branes to the
AdS$_5 \times {\rm S}^5$ background and obtained a model
which had no confinement. A remedy was found shortly.
Flavor in  a  confining background was  introduced in Ref.~\cite{Sakai}
in the context of the 
Klebanov-Strassler model. A simpler and more 
illuminating paper \cite{Kruc} followed almost immediately.
It was based on adding D6 flavor branes to the
model of Witten of near-extremal D4 branes. Although Ref.~\cite{Kruc} 
was very inspiring, the authors themselves were aware of the fact this model did not
incorporate chiral symmetry in the ultraviolet and, hence, 
dynamical chiral symmetry breaking of QCD was not addressed. 

Dedicated designs allowing one to  include  chiral symmetry breaking and related 
low-energy phenomena in AdS/QCD were worked out in \cite{B,E,2,drp}.
In fact, it will be very interesting to check whether fluctuations of the
probe branes that correspond to meson excitations 
will explain chiral symmetry restoration in high excitations, and if yes, 
whether the rate of restoration in highly excited mesons will be properly reproduced.

This last remark presents a nice bridge between the contents of Sect.~\ref{trotcsr}
and the remainder of this talk devoted to implementations of universality.
First AdS/QCD-based analyses of the issue were reported in Refs.~\cite{4,5}.
After a brief review I confront them with QCD expectations
which I derived for this conference. One should note that, 
notwithstanding their stimulating character,
the models \cite{4,5} are not based on backgrounds and
probes  proven  to be  duals of QCD. In fact, in these models
the Regge trajectories are not asymptotically 
linear. Moreover, they share a general feature inherent to all
probe-based constructions. Ignoring back reaction means
that the flavor-carrying quarks in these models are, in essence,
non-dynamical. A conceptual parallel that immediately comes to one's
mind is quenching in lattice QCD. This does not seem to be a serious 
drawback, though, given that $N=\infty$.

One could say that unless the above stumbling blocks 
are eliminated there is no point in analyzing consequences of 
present-day AdS/QCD and confronting them with QCD proper.
Such a  standpoint, although legitimate, is not constructive.
The more aspects we study the more chances we have for 
eliminating drawbacks and finding
a ``perfectly good" holographic dual. 

\subsection{Implementation of universality in AdS/QCD and QCD}

To begin with, I would like to  dwell on the work of Hong et al. \cite{4} 
devoted to the issue  
of VMD and universality of the $\rho$-meson coupling. 

The notion of VMD is known from the 1960's \cite{sakurai}.
Let us consider, for definiteness, the vector isovector current
\beq
J^\mu =\frac{1}{2}\, \left(\bar u\gamma^\mu u -\bar d\gamma^\mu d
\right)\,.
\label{ici}
\eeq
If it is completely saturated by the $\rho$ meson,
\beq
J^\mu \equiv \left( \frac{M_\rho^2}{g_\rho} \right) \rho^\mu\,,
\label{zimo}
\eeq
with no higher excitations, then the $\rho H H $ coupling
is obviously universal and depends only on the isospin of the
hadron $H$.  Indeed, consider the formfactor of $H$ for the 
$J^\mu$-induced transition. At zero momentum 
it equals to the isospin of $H$. On the other hand,
saturating the formfactor by the $\rho$ meson, we get
\beq
\left(g_{\rho H H }\right)\, g_{\rho}^{-1} =H\,\, \mbox{ isospin}\,.
\label{stupid}
\eeq
This is the famous VMD formula. Say, for the pion, $g_{\rho\pi\pi} = g_\rho$.
In what follows I will keep in mind $H=\pi$ as a typical example.\footnote{
In this case the (dimensionless) coupling constant $g_{\rho\pi\pi}$ is defined as
$$\langle\rho|\pi^+\pi^-\rangle = g_{\rho\pi\pi} \, 
\epsilon_\mu \left(p_+^\mu -p_-^\mu\right).$$}

Equation (\ref{stupid}) is approximate since the absolute saturation (\ref{zimo})
is certainly unrealistic. Higher radial excitations are coupled to the isovector current
too,
\beq
\langle\mbox{vac} |J^\mu |\rho_n\rangle = \left( \frac{M_n^2}{g_n} \right)\,\epsilon_n^\mu
\neq 0\,,\qquad n = 1,2,3,....
\eeq
The exact formula  replacing VMD is
\beq
 \left(g_{\rho H H }\right)\, g_{\rho}^{-1} 
 + \sum_{n=1}^\infty \,\frac{g_{\rho_n H H }}{g_n}
 = H\,\, \mbox{ isospin}\,.
 \label{stupidp}
\eeq
If the sum over excitations 
on the left-hand side is numerically small, for whatever reason,
one still recovers the universality relation (\ref{stupid}) which will be valid approximately rather than exactly.

There are two distinct regimes ensuring    suppression of the sum:
(a) each term $n=1,2,3,...$ is individually small; (b) each term 
is of the same order as $ \left(g_{\rho H H }\right)\, g_{\rho}^{-1}$,
but   successive terms are sign-alternating and compensate each other.
The first option is an approximate VMD and, hence, 
leads to a natural universality, while the second one is in fact
an ``accidental" or `` fortuitous" universality.

\begin{figure}[h]
 \centerline{\includegraphics[width=4.5in]{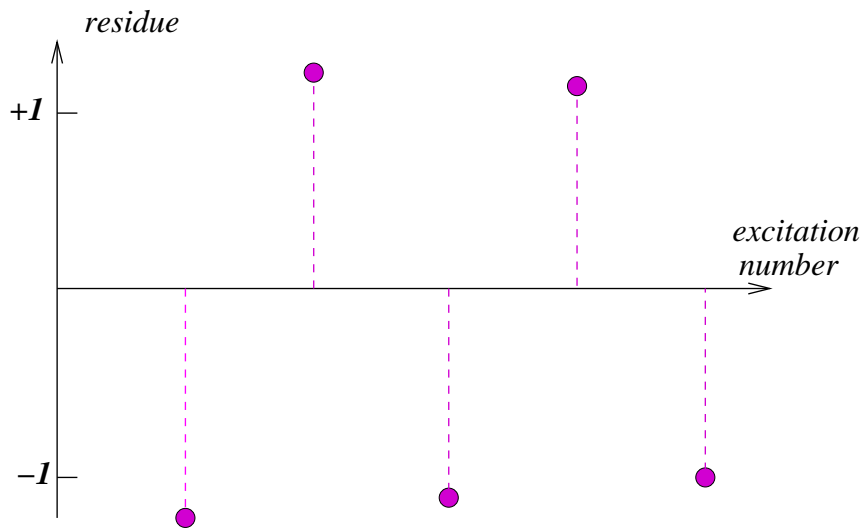}}
 \caption{The residue
$  g_{\rho_n H H}  \, (  g_n)^{-1}$ versus $n$ for $n=1,2,...$ Borrowed from \cite{4}.
 }
 \label{tr2}
 \end{figure}

It is the latter regime which takes place in the 
holographic model considered in Ref.~\cite{4}, 
see Fig.~\ref{tr2} illustrating numerical results 
reported in this paper.\footnote{Whether or not poor convergence
of $ \left(g_{\rho_n H H }\right)\, g_{\rho_n}^{-1}$ is a general feature of AdS/QCD,
or, perhaps,  of a certain class of holographic models,
remains to be seen. According to M. Stephanov's private
communication,  the AdS/QCD model of Ref.~\cite{5},
which builds on previous results \cite{B,E},
yields $n^{-5/2}$,
the rate of fall-off that is even steeper than that expected in QCD, see Sect.
\ref{sir}. An intermediate regime, with the fall-off rate $\sim n^{-1/2}$,
is reported in \cite{vj}. It would be instructive 
to discuss in detail particular reasons explaining the above distinctions.}
For the ground state the authors get
\cite{4} $ \left(g_{\rho H H }\right)\approx1.49\,  g_{\rho}$. 
The factor 1.49 is to be compared with unity in VMD.

I would like to show that   QCD proper gives  the former regime 
rather than the latter ---
suppression of individual contributions of high excitations. The divergence 
between QCD and AdS/QCD must cause no 
surprise since AdS/QCD does not accurately describe 
short-distance QCD dynamics which governs high excitations. 

Now, let us discuss QCD-based expectations in more detail.

\subsection{Sign alternating  residues}
\label{sir}

\begin{figure}[h]
 \centerline{\includegraphics[width=1.4in]{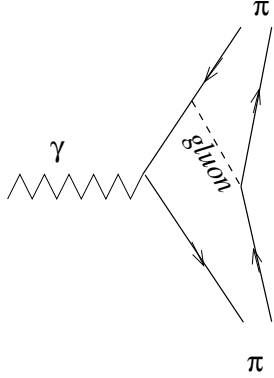}}
 \caption{Asymptotic behavior of the pion formfactor in QCD.
 }
 \label{tr3}
 \end{figure}

As well-known (see e.g. \cite{CZB} for extensive reviews), 
at large (Euclidean)
momentum transfer the pion formfactor is determined by the graph of Fig. \ref{tr3}
and scales as
\beq
F_\pi(Q^2) \sim \frac{1}{Q^2\,\ln Q^2}\,.
\label{tib}
\eeq
It is important that the fall off is {\em faster} than $1/Q^2$.
Comparing Eq.~(\ref{tib}) with the sum-over-resonances representation,
\beq
F_\pi(Q^2) = 
\sum_{n=0}^\infty \,\frac{g_{\rho_n \pi\pi }}{g_n}\,\frac{M^2_{\rho_n }}{Q^2+M^2_{\rho_n }}\,,
\eeq
we immediately conclude that successive terms must be sign-alternating ---
otherwise the asymptotic fall off would be $1/Q^2$.
Thus, QCD supports the sign-alternation feature of AdS/QCD.

\subsection{Large \boldmath{$n$} suppression of the residues}

\begin{figure}[h]
 \centerline{\includegraphics[width=1.4in]{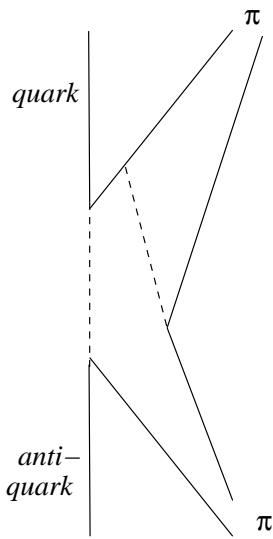}}
 \caption{Determination of $g_{\rho_n \pi\pi }$ 
 at large $n$ from  two-gluon exchange graph.
 }
 \label{tr4}
 \end{figure}

Now, our task is to estimate the large $n$ behavior of 
$g_{\rho_n \pi\pi }/{g_n}$ using, as previously, 
the quasiclassical approximation. A high radial 
excitation of the $\rho$ meson can be viewed as a
an ultrarelativistic quark-antiquark system, each quark having  energy $m_n/2$
Conversion to the pion pair proceeds via the diagram of Fig.~\ref{tr4}.
A straightforward examination of this graph allows us to conclude
that 
\beq
g_{\rho_n \pi\pi } \sim \frac{f_\pi^2}{g_n}\,\frac{1}{M_n^2}\,
\frac{1}{[\ln (M_n^2/\Lambda^2)]^2}\,,
\label{imf}
\eeq
where both factors, $M_n^{-2}$ and the square of the logarithm,
are the consequences of the two-gluon exchange.
The logarithm is due to $\alpha_s$ (see below). 

Deviations from VMD in Eq.~(\ref{stupidp})
are determined by the residues
\beq
\frac{g_{\rho_n \pi\pi }}{g_n} \sim \frac{f_\pi^2}{g_n^2}\,\frac{1}{M_n^2}\,
\frac{1}{[\ln (M_n^2/\Lambda^2)]^2}\,.
\label{imfp}
\eeq
Taking into account 
the fact that $M_n^2/g_n^2$ is asymptotically $n$-independent, 
we arrive at the following suppression factor:
\beq
\left|\frac{g_{\rho_n \pi\pi }}{g_n}\right| \sim \frac{1}{n^2}\, \frac{1}{(\ln n )^2}\,.
\label{imfpp}
\eeq
the fall off of the residues of high excitations is quite steep.
Formally, Eq.~(\ref{imfpp}) is valid at large $n$.
It is reasonable to ask whether a suppression persists for the first or second
excitation. The answer seems to be positive. 
Even if $n\sim 1$, there is a numeric suppression coming from $\alpha_s^2$ of the type
\beq
(\lambda )^2= 
\left( \frac{N\alpha_s}{2\pi} \right)^2\sim \left(4\,\ln M_n/\Lambda \right)^{-2}\,,
\label{ili}
\eeq
where $\lambda$ is the 
't Hooft coupling, small in QCD and large in AdS/QCD.
According to the above estimate, the first excitation contributes 
in Eq.~(\ref{stupidp}) at the 10\% level. This is in accord with the 
experience I gained from multiple analyses of the QCD sum rules \cite{SVZ}
in which I had been involved in the past.

\section{In conclusion...}

I would like to conclude this talk on a curious note showing that, perhaps, indeed,
new is a well-forgotten old. Descending down from conceptual summits
to down-to-earth technicalities let us ask ourselves
what we learn from AdS/QCD with regards, say, to the $\rho$-meson channel
at operational level. To this end let us have a closer look at results
reported in \cite{5}. Operationally, the bare-quark-loop logarithm
is represented in this work as an infinite sum over 
excited $\rho$ mesons
whose masses and residues are adjusted in such a way that
the above infinite sum reproduces pure logarithm of $Q^2$
up to corrections exponentially small at large 
$Q^2$ (there are no power corrections).

A similar question was raised long ago, decades before
AdS/QCD,  by A. Migdal \cite{AAM}, who asked himself
what is the best possible accuracy to which $\log Q^2$
can be approximated by an infinite sum of infinitely narrow
discrete resonances, and what are the 
corresponding values of the resonance masses
and residues. He answered this question as follows:
``the accuracy is exponential at large $Q^2$
and the resonances must be situated at the zeros of a Bessel function."
This is exactly the position of the excited $\rho$ mesons
found in Ref.~\cite{5}\,{!}

\section{Conclusions}

\hspace{0.5cm}$\Diamond\!\!\!\!\star$ \hspace{1mm} 
Chiral symmetry restoration in high radial excitations occurs at the rate
$$|\delta M_n|\sim \Lambda n^{-3/2}$$ or faster
(this is related to the quark condensate $\langle\bar\Psi\Psi\rangle^2$);

\vspace{0.2cm}

$\Diamond\!\!\!\!\star\Diamond\!\!\!\!\star$ \hspace{1mm} 
Relatively simple versions of
AdS/QCD (the majority of the holographic duals
analyzed so far!), along with the wrong $n$ dependence of $M_n$
(the well-known fact) also lead  to a wrong pattern for the 
$n$ dependence of the residue $ g_{\rho_n \pi\pi }/{g_n}$. 
Thus, although the $g_{\rho HH}$
universality is implemented in AdS/QCD, it may be implemented in a wrong way!

\section*{Acknowledgments}

I am grateful to T. Cohen, L. Glozman, T.~Son,  M. Stephanov,
and A. Vainshtein
for illuminating discussions. Very useful 
communications with A. Armoni, S. Beane, M. Chizhov, N. Evans, 
J.~Hirn, D. Jido, I. Kirsch,  J. Sonnenschein, and  M. Strassler are gratefully 
acknowledged. This work was supported in part by DOE grant DE-FG02-94ER408.

\hspace{1cm}


\begin{thebibliography}{99}

\bibitem{K}
  A.~Karch and E.~Katz,
  JHEP {\bf 0206}, 043 (2002)
  [hep-th/0205236]; see also
 A.~Karch, E.~Katz and N.~Weiner,
  Phys.\ Rev.\ Lett.\  {\bf 90}, 091601 (2003)
  [hep-th/0211107].
  
\bibitem{1}
  H.~Boschi-Filho and N.~R.~F.~Braga,
  Eur.\ Phys.\ J.\ C {\bf 32}, 529 (2004)
  [hep-th/0209080].

 \bibitem{Kr}
  M.~Kruczenski, D.~Mateos, R.~C.~Myers and D.~J.~Winters,
  JHEP {\bf 0307}, 049 (2003)
  [hep-th/0304032].
  
  \bibitem{Sakai}
  T.~Sakai and J.~Sonnenschein,
  JHEP {\bf 0309}, 047 (2003)
  [hep-th/0305049].
  
  \bibitem{Kruc}
  M.~Kruczenski, D.~Mateos, R.~C.~Myers and D.~J.~Winters,
  JHEP {\bf 0405}, 041 (2004)
  [hep-th/0311270].
  
  \bibitem{B}
  J.~Babington, J.~Erdmenger, N.~J.~Evans, Z.~Guralnik and I.~Kirsch,
  Phys.\ Rev.\ D {\bf 69}, 066007 (2004)
  [hep-th/0306018].
  
 \bibitem{H} 
S.~Hong, S.~Yoon and M.~J.~Strassler,
  JHEP {\bf 0404}, 046 (2004)
  [hep-th/0312071].
  
  \bibitem{E}
  N.~J.~Evans and J.~P.~Shock,
  Phys.\ Rev.\ D {\bf 70}, 046002 (2004)
  [hep-th/0403279].
  
  \bibitem{Bur}
  B.~A.~Burrington, J.~T.~Liu, L.~A.~Pando Zayas and D.~Vaman,
  JHEP {\bf 0502}, 022 (2005)
  [hep-th/0406207].
  
  \bibitem{Kleba}
  I.~R.~Klebanov and J.~M.~Maldacena,
  Int.\ J.\ Mod.\ Phys.\ A {\bf 19}, 5003 (2004)
  [hep-th/0409133].
  
\bibitem{2}
  T.~Sakai and S.~Sugimoto,
  Prog.\ Theor.\ Phys.\  {\bf 113}, 843 (2005)
  [hep-th/0412141];
{\em More on a holographic dual of QCD,}
  hep-th/0507073.

\bibitem{3}
  G.~F.~de Teramond and S.~J.~Brodsky,
  Phys.\ Rev.\ Lett.\  {\bf 94}, 201601 (2005)
  [hep-th/0501022].
  
\bibitem{4}
 S.~Hong, S.~Yoon and M.~J.~Strassler,
{\em On the couplings of the $\rho$ meson in AdS/QCD},
hep-ph/0501197.

\bibitem{5}
J.~Erlich, E.~Katz, D.~T.~Son and M.~A.~Stephanov,
{\em QCD and a holographic model of hadrons}, 
hep-ph/0501128.
  
\bibitem{drp}
L.~Da Rold and A.~Pomarol,
{\em Chiral symmetry breaking from five-dimensional spaces,}
hep-ph/0501218;
H.~Boschi-Filho, N.~R.~F.~Braga and H.~L.~Carrion,
{\em Glueball Regge trajectories from gauge/string duality and the Pomeron,}
  hep-th/0507063.
  
  \bibitem{Kup}
   M.~Kruczenski, L.~A.~P.~Zayas, J.~Sonnenschein and D.~Vaman,
  JHEP {\bf 0506}, 046 (2005)
  [hep-th/0410035];
  S.~Kuperstein and J.~Sonnenschein,
  JHEP {\bf 0411}, 026 (2004)
  [hep-th/0411009].
  
   \bibitem{Kirsch}
  I.~Kirsch and D.~Vaman,
  Phys.\ Rev.\ D {\bf 72}, 026007 (2005)
  [hep-th/0505164].

\bibitem{ASV}
A.~Armoni, M.~Shifman and G.~Veneziano,
  Phys.\ Rev.\ Lett.\  {\bf 91}, 191601 (2003)
  [hep-th/0307097];
  Phys.\ Lett.\ B {\bf 579}, 384 (2004)
  [hep-th/0309013].

\bibitem{thooft}
G.~'t Hooft,
Nucl.\ Phys.\ B {\bf 72}, 461 (1974).

\bibitem{Gloz0}
L.~Y.~Glozman,
Phys.\ Lett.\ B {\bf 539}, 257 (2002)
[hep-ph/0205072];
T.~D.~Cohen and L.~Y.~Glozman,
Int.\ J.\ Mod.\ Phys.\ A {\bf 17}, 1327 (2002)
[hep-ph/0201242];

\bibitem{Gloz1}
L.~Y.~Glozman,
AIP Conf.\ Proc.\  {\bf 717}, 726 (2004)
[hep-ph/0309334].

\bibitem{Gloz2}
L.~Y.~Glozman,
Phys.\ Lett.\ B {\bf 587}, 69 (2004)
[hep-ph/0312354].

\bibitem{Gloz3}
L.~Y.~Glozman,
{\em Restoration of chiral symmetry in excited hadrons,}
Lectures at 44-th Cracow School of Theoretical Physics {\sl New Results in Particle Physics}, Zakopane, Poland,  May   2004,
hep-ph/0410194; this is a review where
the reader can find references  to earlier works.

\bibitem{NSVZ}
V.~A.~Novikov, M.~A.~Shifman, A.~I.~Vainshtein and V.~I.~Zakharov,
  Nucl.\ Phys.\ B {\bf 191}, 301 (1981);
M.~A.~Shifman,
  Sov.\ J.\ Nucl.\ Phys.\  {\bf 36}, 749 (1982)
  [Yad.\ Fiz.\  {\bf 36}, 1290 (1982)].
  
\bibitem{Gloz4}
L.~Y.~Glozman,
{\em Chiral and U$(1)_A$ restorations high in the hadron spectrum, semiclassical  approximation and large $N_c$,}
hep-ph/0411281.

\bibitem{jido}
A review can be found e.g. in
 T.~D.~Cohen and X.~D.~Ji,
  Phys.\ Rev.\ D {\bf 55}, 6870 (1997)
  [hep-ph/9612302]. For a  discussion
  of SU(2)$\times$SU(2) representations for baryons
  unrelated to dynamical  issues relevant to  high excitations
   the reader is referred to
 D.~Jido, T.~Hatsuda and T.~Kunihiro,
  Phys.\ Rev.\ Lett.\  {\bf 84}, 3252 (2000)
  [hep-ph/9910375]; 
D.~Jido, M.~Oka and A.~Hosaka,
  Prog.\ Theor.\ Phys.\  {\bf 106}, 873 (2001)
  [hep-ph/0110005].

\bibitem{Ch}
 M.~V.~Chizhov,
{\em Tensor excitations in Nambu--Jona-Lasinio model,}
 hep-ph/9610220; 
JETP Lett.\  {\bf 80}, 73 (2004)
   [hep-ph/0307100].

\bibitem{TM}
G. 't Hooft, {\it Nucl. Phys.} {\bf B75} (1974) 461 [Reprinted in
G. 't Hooft, {\em Under the Spell of the Gauge Principle} (World 
Scientific, Singapore 1994), page 443]; see also F. Lenz, M. Thies, S. 
Levit and K. Yazaki, 
{\it Ann.  Phys.} (N.Y.) {\bf 208} (1991) 1;
C. Callan, N. Coote, and D. Gross,
{\it Phys. Rev.} {\bf D13} (1976) 1649;
M. Einhorn, {\it Phys. Rev.} {\bf D14} (1976) 3451;
M. Einhorn, S. Nussinov, and E. Rabinovici, 
 {\it Phys. Rev.} {\bf D15} (1977) 2282.
 
\bibitem{TM1}
I. Bars and M. Green, {\it Phys. Rev.} {\bf D17} (1978) 537.

\bibitem{Nussinov} 
A.~Casher, H.~Neuberger and S.~Nussinov,
  Phys.\ Rev.\ D {\bf 20}, 179 (1979).

\bibitem{BS}
B.~Blok, M.~A.~Shifman and D.~X.~Zhang,
Phys.\ Rev.\ D {\bf 57}, 2691 (1998); (E)
 D {\bf 59}, 019901 (1999)
[hep-ph/9709333].

\bibitem{Schreiber}
  E.~Schreiber,
 {\em Excited mesons and quantization of string endpoints,}
  hep-th/0403226.

\bibitem{SVZ}
M.~A.~Shifman, A.~I.~Vainshtein and V.~I.~Zakharov,
  Nucl.\ Phys.\ B {\bf 147}, 385; 448 (1979).
   
  \bibitem{BBB}
   S.~R.~Beane,
  Phys.\ Rev.\ D {\bf 64}, 116010 (2001)
  [hep-ph/0106022].
   
   \bibitem{AAA}
 S.~S.~Afonin, A.~A.~Andrianov, V.~A.~Andrianov and D.~Espriu,
  JHEP {\bf 0404}, 039 (2004)
  [hep-ph/0403268].
   
\bibitem{AdSCFT}
J.~M.~Maldacena,
Adv.\ Theor.\ Math.\ Phys.\  {\bf 2}, 231 (1998)
[Int.\ J.\ Theor.\ Phys.\  {\bf 38}, 1113 (1999)]
[hep-th/9711200];
S.~S.~Gubser, I.~R.~Klebanov and A.~M.~Polyakov,
Phys.\ Lett.\ B {\bf 428}, 105 (1998)
[hep-th/9802109];
E.~Witten,
Adv.\ Theor.\ Math.\ Phys.\  {\bf 2}, 253 (1998)
[hep-th/9802150].
   
\bibitem{sakurai}  
J.J. Sakurai, {\sl Currents and Mesons},
(Univ. of Chicago Press, 1969).
 
 \bibitem{vj}
 J.~Hirn and V.~Sanz,
  {\em Interpolating between low and high energy QCD via a 5D Yang--Mills model,}
  hep-ph/0507049.
 
  \bibitem{CZB}
V.~L.~Chernyak and A.~R.~Zhitnitsky,
{\em Asymptotic Behavior Of Exclusive Processes In QCD,}
Phys.\ Rept.\  {\bf 112}, 173 (1984);
S.~J.~Brodsky and G.~P.~Lepage,
{\em Exclusive Processes In Quantum Chromodynamics,}
Adv.\ Ser.\ Direct.\ High Energy Phys.\  {\bf 5}, 93 (1989)
[reprinted in {\sl  Perturbative Quantum Chromodynamics}, Ed. A.H. Mueller (World Scientific, Singapore, 1989), p. 93].

\bibitem{AAM}
 A.~A.~Migdal,
  Annals Phys.\  {\bf 110}, 46 (1978);
H.~G.~Dosch, J.~Kripfganz and M.~G.~Schmidt,
  Phys.\ Lett.\ B {\bf 70}, 337 (1977).




  
\end{thebibliography}
\end{document}